\documentclass[aps,prb,preprint]{revtex4-1}
\usepackage{graphicx}

\usepackage{bm}

\begin{document}


\title{Natural optical activity and its control by electric field in electrotoroidic systems}

\author{Sergey Prosandeev$^{1,2}$, Andrei Malashevich$^{3}$, Zhigang Gui$^{1}$, Lydie Louis$^{1}$, Raymond Walter$^{1}$, Ivo Souza$^{4}$ and L. Bellaiche$^{1}$}

\affiliation{$^{1}$Physics Department,University of Arkansas, Fayetteville, Arkansas 72701, USA\\
  $^2$Institute of Physics, South Federal University, Rostov on Don, Russia, 344090 \\
  $^{3}$Department of Applied Physics, Yale University, New Haven, Connecticut
06511, USA\\ $^{4}$Centro de F\'{\i}sica de Materiales (CSIC) and DIPC,
  Universidad del Pa\'is Vasco, 20018 San
  Sebasti\'an, and Ikerbasque Foundation, 48011 Bilbao, Spain}

\date{\today}

\begin{abstract}
  We propose the existence, via analytical derivations, novel phenomenologies, and
  first-principles-based simulations, of a
 new class of materials that are not only
  spontaneously optically active, but also for which the sense of rotation can be switched by an
  electric field applied to them-- via an induced transition between the
  dextrorotatory and laevorotatory forms. Such systems possess electric vortices that are
  coupled to a spontaneous electrical polarization. Furthermore, our
  atomistic simulations provide a deep microscopic insight into, and
  understanding of, this class of naturally optically active
  materials.
\end{abstract}


\maketitle

\vspace{2mm}

\section{Introduction}

The speed of propagation of
  circularly-polarized light traveling inside an {\it optically active} material
 depends on its helicity \cite{Melrose,Barron}.  Accordingly, the plane of
  polarization of linearly polarized light rotates by a fixed amount
  per unit length, a phenomenon known as {\it optical rotation}.
One traditional way to make materials
optically active is to take advantage of the Faraday effect, by
applying a magnetic field. However, there are some specific systems
that are {\it naturally gyrotropic}, that is they spontaneously
possess optical activity. Examples of known natural gyrotropic systems are
quartz \cite{Arago}, some organic liquids and aqueous solutions of
sugar and tartaric acid \cite{Melrose}, the Pb$_5$Ge$_3$O$_{11}$
compound \cite{experiment,Koiiak}, and the layered crystal
(C$_5$H$_{11}$NH$_3$)$_2$ZnCl$_4$ \cite{Pnma-P212121}. Finding novel
natural gyrotropic materials has great
fundamental interest. It may also lead to the design of
novel devices, such as optical circulators and
amplifiers, especially if the {\it sign} of the optical rotation can be
efficiently controlled by an external factor that is easy to
manipulate.

When searching for new natural gyrotropic
materials, one should remember the observation of Pasteur that
 chiral crystals display spontaneous optical activity, which reverses sign
 when going from the original structure to its mirror image \cite{Pasteur}.
 Hence it is worthwhile to consider
a newly discovered class of materials that are potentially chiral, and
therefore may be naturally gyrotropic. This class is formed by
electrotoroidic compounds (also called ferrotoroidics
\cite{Shmid}). These are systems that possess an
electrical toroidal moment, or equivalently, exhibit electric vortices
\cite{Dubovik}. Such intriguing compounds were predicted to exist around
nine years ago \cite{Ivan}, and were found
  experimentally only recently
\cite{Gruverman-Scott-vortex,Balke-Kalinin,Vasudevan,Nelson,Gregg}.
One may therefore wonder if this new class of materials is indeed
naturally gyrotropic, and/or if there are other necessary
conditions, in addition to the existence of an electrical
toroidal moment, for such materials to be optically active.

In this work, we carry out
analytical derivations, original phenomenologies and first-principles-based
computations that
successfully address all the aforementioned important issues. In
particular, we find that electrotoroidic materials
do possess spontaneous optical activity, but only if their electric
toroidal moment changes {\it linearly} under an applied electric field. This linear
dependence is further proved to occur if the
electrotoroidic materials also possess a spontaneous
electrical polarization that is coupled to the electric toroidal
moment, or if they are also piezoelectric with the strain affecting
the value of the electric toroidal moment. We also find that, in the former case,
the applied electric field further allows the control of the sign of the optical activity.
Our atomistic approach also reveals the evolution of the
microstructure leading to the occurrence of field-switchable
gyrotropy, and  shows that the optical rotatory
  strength can be significant in some
electrotoroidic systems.



\section{Relation between gyrotropy and electrical toroidal moment in electrotoroidic systems}

Let us first recall that the gyrotropy tensor elements, $g_{ml}$, are
defined via \cite{Landau}:
\begin{equation}
\label{eq74}
g_{mk}=\frac{\omega }{2c}e_{ijm}\gamma _{ijk}
\end{equation}
where $e_{ijm}$ is the Levi-Civita tensor
\cite{Tyldesley}, $c$ is the speed of light, and $\omega$ is the angular
frequency. Note that this angular frequency  is not restricted to the optical range. For instance, it can also correspond to the 1-100 GHz frequency range.
The $\gamma$ tensor provides the linear dependence of
the dielectric permittivity on the wave vector ${\bf k}$ in the
optically active material, that is:
\begin{equation}
\label{eq15}
\varepsilon _{ik} \left( {\omega ,{\rm {\bf k}}} \right)=\varepsilon
_{ik}^{\left( 0 \right)} \left( \omega \right)+i\gamma _{ikl} k_l
\end{equation}
Here, $k_l$ is the $l$-component of the wave vector;
$\varepsilon _{ik} \left( {\omega ,{\rm {\bf k}}} \right)$ denotes the
double Fourier transform in time and space of the dielectric tensor, with the
long-wavelength components being
denoted by $\varepsilon_{ik}^{\left( 0 \right)}$. Throughout this
manuscript we adopt Einstein notation, in which one implicitly sums
over repeated indices (as it happens, e.g., for the $l$ index in
Eq. (\ref{eq15})).

Thus, the calculation of the gyrotropy tensor can be reduced to the calculation of the tensor $\gamma$ responsible for the spatial dispersion of the dielectric permittivity.

Alternatively, one can use the following formula for the dielectric
permittivity \cite{Landau,Melrose}:
\begin{equation}
\label{eq19}
\varepsilon _{ik}\left( {\omega ,{\rm {\bf k}}} \right) \\
=\delta _{ik} +\frac{4\pi i}{\omega }\sigma _{ik}\left( {\omega ,{\rm {\bf k}}} \right) \\
=\delta _{ik} +\frac{4\pi i}{\omega }(\sigma _{ik}^{(0)}\left( {\omega } \right)+\sigma_{ikl}k_l)
\end{equation}
where $\delta _{ik}$ is the Kronecker symbol and $\sigma _{ik}\left( {\omega ,{\rm {\bf k}}} \right)$ is the  effective conductivity tensor in the reciprocal space, at a given frequency \cite{Melrose}. $\sigma_{ikl}$ is the third-rank tensor associated with the linear dependence of the effective conductivity tensor on the wave vector, and $\sigma _{ik}^{(0)}$ is the effective conductivity tensor at zero wave vector.
Combining Eqs. (\ref{eq19})  with Eq. (\ref{eq15}) yields:
\begin{equation}
\label{eq20}
\gamma _{ikl} =\frac{4\pi}{\omega }\sigma_{ikl}\\
=\frac{4\pi}{\omega }(\sigma_{ikl}^S(\omega)+\sigma _{ikl}^A(\omega))
\end{equation}
where
\begin{equation}
\label{eq16b}
\sigma_{ijk}^A =\frac{1}{2}(\sigma_{ijk}-\sigma_{jik})
\end{equation}
and
\begin{equation}
\label{eq16c}
\sigma _{ijk}^S =\frac{1}{2}(\sigma_{ijk}+\sigma_{jik})
\end{equation}
Moreover, using the results of Ref. \cite{Malashevich} and working at nonabsorbing frequencies (i.e., frequencies, such as GHz in ferroelectrics, for which the corresponding energy is below the band gap of the material), one can write

\begin{equation}
\label{eq16}
\sigma _{ijk}^A =ic\left( {e_{jkl} \beta _{il} -e_{ikl} \beta _{jl} }
\right)+\omega \xi _{ijk}
\end{equation}
with
\begin{equation}
\label{eq17}
\beta _{ij} =i\mathrm{Im}(\chi _{ij}^{em}) \mathop = -i\mathrm{Im}(\chi _{ji}^{me})
\end{equation}
and
\begin{equation}
\label{eq18}
\xi _{ijk} =\frac{1}{2}\left[ {\frac{dQ_{kj} }{dE_i }-\frac{dQ_{ki} }{dE_j
}} \right]
\end{equation}
where $\mathrm{Im}$ stands for the imaginary part and $Q$ is the quadrupole moment of the system \cite{Raab}. $\chi^{me}$ is the response of the magnetization, {\bf M}, to an electric field {\bf E}, while $\chi^{em}$ is the response of the electrical polarization, {\bf P}, to a magnetic field {\bf B}, that is:
\begin{equation}
\label{eq22}
\chi _{ij}^{me} =\frac{dM_i }{dE_j } \\ \nonumber ~~~{\rm and} ~~~
\chi _{ji}^{em} =\frac{dP_j }{dB_i}
\end{equation}

Inserting Eq. (\ref{eq16}) into Eq. (\ref{eq20}) provides :
\begin{equation}
\label{eq21}
\gamma _{ijk} =\frac{4\pi }{\omega }\left[ {c\left( {e_{jkl} \mathrm{Im}\,\chi
_{li}^{me} -e_{ikl} \mathrm{Im}\,\chi _{lj}^{me} } \right)+\omega \xi _{ijk} } \right] + \gamma_{ijk}^S
\end{equation}
where $\gamma_{ijk}^S=(4\pi/\omega)\sigma_{ijk}^S$ is the contribution of the symmetric part of the conductivity to the $\gamma$ tensor. As a result, $\gamma_{ijk}^S$ is non-zero only when the system is magnetized or possesses a spontaneous magnetic order \cite{Landau}.

Let us now focus on the magnetization, which can be written as
 \cite{Raab}:
\begin{equation}
\label{eqm1}
{\bf M} =\frac{1}{2cV} \int {\left ( {\bf r} \times {\bf {\cal J}(r)}\right ) d^3r }
\end{equation}
where $c$ is the speed of light, $V$ is the volume of the system, {\bf
  r} is the position vector, and ${\bf {\cal J}(r)}$ is the current density. We
consider here the following contributions to this density:
\begin{equation}
\label{eq6}
{\bf {\cal J}(r)}=\dot {{\cal P}}({\bf r)}+c~ {\bm \nabla}\times {\bf {\cal M}_0(r)}
\end{equation}
where the dot symbol refers to the partial derivative with respect to time. ${\bf {\cal P}(r)}$  is the polarization {\it field}, that is, the quantity for which the spatial average is the macroscopic polarization. Similarly, ${\bf {\cal M}_0(r)}$ is the magnetization field, that is, the quantity
for which the spatial average is the part of the macroscopic magnetization that does not originate
from the time derivative of the polarization field \cite{gauge}.
By plugging this latter equality into Eq. (\ref{eqm1}), we have:
\begin{equation}
\label{eqn6}
{\bf M} =\frac{1}{2cV} \int{ \left ( {\bf r} \times  \dot {{\cal P}}({\bf r)}\right ) d^3r}+
\frac{1}{2V} \int{ \left ( {\bf r} \times  {\bm \nabla}\times {\bf {\cal M}_0(r)} \right ) d^3r}
=\frac{1}{2cV} \int{ \left ( {\bf r} \times \dot {{\cal P}}({\bf r)}\right ) d^3r} + {\bf M}_0
\end{equation}
The analytical expression
of this latter equation bears some similarities with the definition of
the electrical toroidal moment, {\bf G}, that is \cite{Dubovik}
\begin{equation}
\label{eqn1}
{\bf G}=\frac{1}{2V}\int {\left ( {\bf r} \times {\bf {\cal P}(r)}  \right ) d^3r}~~~,
\end{equation}

More precisely, taking the time derivative of {\bf G} gives:
\begin{equation}
\label{eqntds}
\dot{\bf G} \simeq \frac{1}{2V} \int{ \left ( {\bf r} \times \dot {{\cal P}}({\bf r)}\right ) d^3r}
\end{equation}
when omitting the time dependency of the volume  (the numerical simulations presented below indeed show that one can safely neglect this dependency  when computing the time derivative of the electric toroidal moment).

As a result, combining Eq. (\ref{eqntds})  and Eq. (\ref{eqn6})  for a monochromatic wave having an $\omega$ angular frequency gives:
\begin{equation}
\label{eqn3}
{\bf M}-{\bf M}_0 \simeq \frac{1}{c}{\dot{\bf G}}=-\frac{i\omega }{c}{\bf G}
\end{equation}
in electrotoroidic systems.

Plugging this latter equation in  Eq. (\ref{eq22})  then gives:
\begin{equation}
\label{eq24}
\chi _{ij}^{me} =\chi_{ij}^{me(0)}-\frac{i\omega }{c}\frac{dG_i }{dE_j }
\end{equation}
where $\chi_{ij}^{me(0)}$ is the magnetoelectric tensor related to the derivative of
${\bf M}_0$ with respect to an electric field. Therefore
\begin{equation}
\label{eq25}
\mathrm{Im}\,(\chi _{ij}^{me}-\chi_{ij}^{me(0)}) =-\frac{\omega }{c}\frac{dG_i }{dE_j }
\end{equation}
This relation between the
  imaginary part of the magnetoelectric susceptibility and the
  field derivative of the electrical toroidal moment is reminiscent of
  the connection discussed in Ref.~\cite{Spaldin} between the linear
  magnetoelectric response and the {\it magnetic} toroidal moment.

Inserting  Eqs. (\ref{eq25}) and (\ref{eq18}) into  Eq. (\ref{eq21}) then provides:
\begin{eqnarray}
\label{eq29}
\gamma _{ijk} =&&\gamma_{ijk}^{S} +\frac{4\pi c}{\omega} \left ( e_{jkl} {\rm Im} \chi_{li}^{me(0)}-e_{ikl}{\rm Im} \chi_{lj}^{me(0)} \right )  \nonumber \\
&&+4\pi \left[ {e_{ikl} \frac{dG_l }{dE_j }-e_{jkl} \frac{dG_l
}{dE_i }+\frac{1}{2}\left( {\frac{dQ_{kj} }{dE_i }-\frac{dQ_{ki} }{dE_j }}
\right)} \right]
\end{eqnarray}

Combining this latter equation with Eq. (\ref{eq74}), and recalling that $\gamma_{ijk}^{S}$ is a symmetric tensor while $e_{ijm}$ is antisymmetric (which makes their product vanishing), gives:
\begin{eqnarray}
\label{eq33before}
g_{mk} =&&4\pi \left ( \delta_{mk} {\rm Im} \chi_{ll}^{me(0)}-{\rm Im}\chi_{mk}^{me(0)}  \right )  \nonumber \\
&& +\frac{4\pi \omega }{c}\left[ \left( {\frac{dG_m }{dE_k }-\frac{dG_l }{dE_l
}\delta _{mk} } \right) +\frac{1}{4} e_{ijm}\left( \frac{dQ_{kj}}{dE_i}-\frac{dQ_{ki}}{dE_j} \right) \right]
\end{eqnarray}

Choosing a specific gauge \cite{gauge} and neglecting quadrupole moments
(simulations reported below show that spontaneous and field-induced quadrupole moments can be neglected for the ferrotoroidics numerically studied in Section IV) lead to the reduction of Eq. (\ref{eq33before}) to:
\begin{equation}
\label{eq33}
g_{mk} =\frac{4\pi \omega }{c}\left[ \left( {\frac{dG_m }{dE_k }-\frac{dG_l }{dE_l
}\delta _{mk} } \right)  \right]
\end{equation}

This  formula nicely reveals that optical activity should happen when electrical toroidal moment  {\it linearly} responds to an applied electric field.

\section{Necessary conditions for gyrotropy in electrotoroidic systems}

According to Eq. (\ref{eq33}), an electrotoroidic system possessing non-vanishing derivatives of its electrical toroidal moment with respect to the electric field automatically possesses natural optical activity. Let us now prove analytically  that the occurrence
of such non-vanishing derivatives requires additional symmetry
breaking in electrotoroidic systems, namely that an electrical
polarization or/and piezoelectricity should also exist,
as well as couplings between electrical toroidal moment and
electric polarization and/or strain.

 For that, let us express the free energy
of an electrotoroidic system that exhibits couplings between
electrical toroidal moment \textbf{G}, polarization \textbf{P}, and
strain $\eta$ as:
\begin{equation}
\label{eq34}
F=F_0+\zeta_{ijkl}G_iG_j\eta_{kl}+\lambda_{ijkl}G_iG_jP_kP_l+q_{ijkl}P_iP_j\eta_{kl}-h_iG_i
\end{equation}
where $h_i=( {\bm \nabla}\times \,{\bf E})_i$ is the field conjugate of $G_i$.

The equilibrium condition, $\partial F / \partial G_n = 0$, implies that
\begin{equation}
\label{eq35}
\partial F_0 / \partial G_n+(\zeta_{njkl}+\zeta_{jnkl})G_j\eta_{kl}+(\lambda_{njkl}+\lambda_{jnkl})G_jP_kP_l
=h_n
\end{equation}
which indicates that $h_n$ depends on both the polarization and strain.

As a result, the change in electrical toroidal
moment with electric field can be separated
into the following two contributions:

\begin{equation}
\label{eq34b}
\frac{d{\rm { G_i}}}{d{\rm { E_j}}}= \left( {\frac{dG_i }{dE_j }} \right)^{(1)}+ \left( {\frac{dG_i }{dE_j }} \right)^{(2)}
\end{equation}
with
\begin{equation}
\label{eq35a}
\left( \frac{dG_i }{dE_j } \right)^{(1)}=
 \frac{dG_i}{d h_n}  \frac{\partial h_n}{\partial  P_l}  \frac{d P_l}{d E_j} = \chi^{(G)}_{in}\frac{\partial h_n}{\partial
P_l}\chi^{(P)}_{lj}
\end{equation}
and
\begin{equation}
\label{eq35b}
\left( {\frac{dG_i }{dE_j }} \right)^{(2)}= \frac{d G_i}{\partial h_n} \frac{\partial h_n}{\partial \eta_{kl}} \frac{d \eta_{kl}}{d E_j} = \chi^{(G)}_{in}
\frac{\partial h_n}{\partial \eta_{kl} } d_{klj}
\end{equation}
Here
\begin{equation}
\chi^{(G)}_{in}=\frac{dG_i}{dh_n}
\end{equation}
is the response of the electrical toroidal moment to its conjugate
field,
\begin{equation}
\chi_{ij} ^{(P)}=\frac{dP_i}{dE_j}
\end{equation}
is the electric susceptibility, and
\begin{equation}
d_{ijk}=\frac{d \eta_{ij}}{dE_k}
\end{equation}
is a piezoelectric tensor.

The remaining derivatives appearing in Eqs. (\ref{eq35a}) and (\ref{eq35b})
can be found from Eq. (\ref{eq35}):
 \begin{equation}
\label{eq38}
\left( {\frac{\partial h_n }{\partial P_l }} \right)= (\lambda_{njlm} +\lambda _{njml}+\lambda_{jnlm} +\lambda _{jnml})G_jP_m
\end{equation}
and
\begin{equation}
\label{eq39}
\left( {\frac{\partial h_n }{\partial \eta_{kl} }} \right)=(\zeta_{njkl}+\zeta _{jnkl}) G_j
\end{equation}

Equations (\ref{eq34b})-(\ref{eq39}) reveal that there are two scenario for the
  occurence of natural optical activity in electrotoroidic systems.
   In the first scenario, the
system possesses a finite polarization that has a bilinear coupling
with the electrical toroidal moment (see Eqs. (\ref{eq35a}), (\ref{eq38}), and (\ref{eq34})). In
the second scenario, the electrotoroidic system is also piezoelectric, and
electrical toroidal moment and strain are coupled to each other (see
Eqs. (\ref{eq35b}), (\ref{eq39}), and (\ref{eq34})). An example of the latter can be found in
Reference \cite{Prosandeev}, where a pure gyrotropic phase transition
leading to a piezoelectric, but non-polar, $P2_12_12_1$ state (that
exhibits spontaneous electrical toroidal moments) was discovered in a
perovskite film. Next, we describe the theoretical prediction of
  a material where the former scenario is realized.

\section{Prediction and microscopic understanding of
gyrotropy in electrotoroidic systems}

The system we have investigated numerically is
a nanocomposite made of periodic squared arrays of BaTiO$_3$ nanowires
embedded in a matrix formed by (Ba,Sr)TiO$_3$ solid solutions having a
85\% Sr composition. The nanowires have a long axis
oriented along the [001] pseudo-cubic direction (chosen to be the
$z$-axis). They possess a squared cross-section of 4.8x4.8 nm$^2$
 in the ($x$,$y$) plane, where the $x$- and $y$-axes are chosen
along the pseudo-cubic [100] and [010] directions, respectively. The
 distance (along the $x$- or $y$-directions) between
 adjacent BaTiO$_3$ nanowires is  2.4 nm.

 We choose this particular nanocomposite system because a recent theoretical study \cite{submitted},
using an effective Hamiltonian ($H_{\rm{eff}}$) scheme, revealed that its ground state possesses a
spontaneous polarization along the $z$-direction inside the whole
composite system, as well as  electric vortices in the ($x$,$y$) planes inside each BaTiO$_3$ nanowire, with the same sense of vortex rotation in every wire. Such a
phase-locking, ferrotoroidic and polar state is shown in Fig. 1a. It
exhibits an electrical toroidal moment being parallel to the polarization.  Figure 1a
also reveals the presence of antivortices
located in the {\it medium}, half-way between the centers of adjacent
vortices.

In the present study, we use the same $H_{\rm{eff}}$ as in Ref. \cite{submitted}, combined  with molecular dynamics techniques, to determine the response of this peculiar state to an $ac$ electric field applied along the main, $z$-direction of the wires. In our simulations, the amplitude of the field was fixed
at 10$^9$ V/m and its frequency ranged between 1GHz and 100GHz. The sinusoidal frequency-driven variation of the electric field with time makes therefore this field ranging in time between
10$^9$ V/m (field along [001]) and -10$^9$ V/m (field along [00-1]).
The idea here is to
check if the electrical toroidal moment has a {\it linear} variation with this field at these investigated frequencies, and therefore if the investigated system can possess nonzero gyrotropy coefficients (see Eq.(\ref{eq33})).

In this effective Hamiltonian method, developed in Ref.\cite{Walizer2006} for (Ba,Sr)TiO$_3$ (BST) compounds, the degrees of freedom are:  the local mode vectors in each 5-atom unit cell (these local modes are directly proportional to the electric dipoles in these cells), the homogeneous strain tensor and inhomogeneous-strain-related variables \cite{Zhong1995}. The total internal energy contains a local mode self-energy, short-range and long-range interactions between local modes, an elastic energy and interactions between local modes and strains. Further energetic terms model the effect of the interfaces between the wires and the medium on electric dipoles and strains, as well as take into account the strain that is induced by the size difference between Ba and Sr ions and its effect on physical properties. The parameters entering the total internal energy are derived from first principles. This $H_{\rm{eff}}$ can be used within Monte-Carlo or Molecular dynamics simulations to obtain finite-temperature static or dynamical properties, respectively, of relatively large supercells (i.e., of the order of thousands or tens of thousands of atoms).  Previous calculations \cite{Choudhury2011,Walizer2006,Lisenkov2007,Hlinka2008,Quingteng2010} for various disordered or ordered BST systems demonstrated the accuracy of this method for several properties. For instance, Curie temperatures and phase diagrams, as well as the subtle temperature-gradient-induced polarization, were well reproduced in BST materials. Similarly, the existence of two modes (rather than a single one as previously believed for a long time) contributing to the GHz-THz dielectric response of pure BaTiO$_3$ and disordered BST solid solutions were predicted via this numerical tool and experimentally confirmed.

Figures 2(a) and 2(b) report the evolution of the $z$-component of the electrical toroidal moment, G$_z$,  and of the polarization, P$_z$, respectively, as a function of the electric field, for a frequency of 1GHz at a temperature of 15K. In practice,  G$_z$ is  computed within a lattice model
\cite{submitted}, by summing over the electric dipoles located at the lattice sites rather than by continuously integrating the polarization field of Eq. (\ref{eqn1}) over the space occupied by the nanowires. The panels in Fig. 1 show snapshots of important states occurring during these hysteresis loops, in order to understand gyrotropy at a microscopic level.
A striking piece of information revealed by Fig. 2 (a) is that G$_z$
{\it linearly decreases} with a slope of $-1.6$ e/V when the applied
$ac$ field varies between 0 (state 1) and its maximum value of 10$^9$
{\rm V/m} (state 2). Such variation therefore results in {\it
  positive} g$_{11}$ and g$_{22}$ gyrotropy coefficients that are both
equal to $0.94 \times 10^{-7}$ for a frequency of 1GHz, according to
Eq. (\ref{eq33}) (that reduces here to $g_{11} =g_{22}=-\frac{\omega
}{c\varepsilon_0} {\frac{dG_z }{dE}}$ in S.I. units, since there are
no $x$- and $y$-components of the toroidal moment and since the field
is applied along $z$ in the studied case).  Interestingly, we
found that the aforementioned slope of $-1.6$
e/V stays roughly constant over the entire frequency
  range we have investigated (up to 100GHz). As a
result, Eq. (\ref{eq33}) indicates that $g_{11} =g_{22}$ should be
proportional to the angular frequency $\omega$ of the applied $ac$
field, and that the meaningful quantity to consider here is the ratio
between $g_{11}$ and this frequency. Such ratio is presently equal to
$5.9 \times 10^{-16}$  per Hz.  Moreover, the rate of optical rotation is related to the
product between $\omega/c$ and the  gyrotropy coefficient according to Ref. \cite{Landau}. As a result, the rate of optical rotation depends on the {\it square} of the angular frequency because of
Eq. (\ref{eq33}), as consistent with one finding of Biot in 1812 \cite{Barron}.
Here, the ratio of the rate of optical rotation
to the square of the angular frequency is
found  to be four orders of magnitude larger than that measured in
``typical'' gyrotropic materials, such as Pb$_5$Ge$_3$O$_{11}$
\cite{experiment,Koiiak}.
As a result, the plane of
  polarization of light will rotate by around $1.2$ radians per meter at
100GHz (or by $1.24\times 10^{-4}$ radians per meter at 1GHz), when
passing through the system.

Figure 2b  indicates that the observed decrease of G$_z$  is accompanied by an increase of the polarization, which is consistent with our numerical finding that increasing the field from 0 to 10$^9$ V/m reduces the $x$- and $y$-components of the electric dipoles inside the nanowires (that form the vortices) while enhancing the $z$-component of the electric dipoles in the whole nanocomposite (i.e., wires and medium). Interestingly, the antivortices in the medium progressively disappear during this linear decrease of G$_z$ and increase of P$_z$, as shown in Figs 1.
Figures 2 also show that decreasing the electric field from 10$^9$
V/m (state 2) to $\simeq$ -0.031 $\times$ 10$^9$  V/m (state 3) leads
to a linear increase of the electric toroidal moment (yielding the
aforementioned values of $g_{11}$ and $g_{22}$), while the
$z$-component of the polarization decreases but still stay
positive.

Further increasing the magnitude of negative electric fields
up to $\simeq$ -0.094 $\times$ 10$^9$ V/m results in drastic changes
for the microstructure: dipoles in the medium now adopt negative
$z$-components (state 3), then sites at the interfaces between
the medium and the wires also flip the sign of the $z$-component of
their dipoles (states 3 and $\alpha$). During these changes, the
overall polarization rapidly varies from a significant positive value
along the $z$-axis to a slightly negative value (Fig. 2b), while G$_z$
is nearly constant, therefore rendering the gyrotropic coefficients
null. Then, continually increasing the strength of the negative $ac$
field up to $\simeq$ -0.48 $\times$ 10$^9$ V/m leads to the next stage:
dipoles {\it inside} the wires begin to change the sign of their
$z$-components (states $\beta$, 4 and $\gamma$) until all of the $z$-components of
these dipoles point down (state 5). During that process, P$_z$ becomes
more and more negative, while the electrical toroidal moment decreases
very fast but remains positive (indicating that the chirality of the
wires is unaffected by the switching of the overall polarization).

Once this process is completed, further increasing the magnitude of
the applied field along [00$\bar{1}$] up to -10$^9$ V/m (state 2$'$),
leads to a {\it linear decrease} of the electrical toroidal
moment. Interestingly, this decrease is quantified by a slope
$dG_z/dE$ that is exactly {\it opposite} to the corresponding one when
going from state 1 to state 2. As a result, the $g_{11}$ and $g_{22}$
gyrotropic coefficients associated with the evolution from state 5 to
state 2$'$ are now {\it negative} and equal $-$$0.94 \times
10^{-7}$ at 1GHz.

Finally, Figures 1 and 2 indicate that varying now
the $ac$ field from its minimal value of -10$^9$ V/m to its maximal
value of 10$^9$ V/m leads to the following succession of states: 2$'$,
5, 1$'$, 3$'$, $\alpha$$'$, $\beta$$'$, 4$'$, $\gamma$$'$, 5$'$ and 2,
where the $'$ superscript used to denote the i$'$ states (with $i$=2, 3,
4, 5, $\alpha$, $\beta$ and $\gamma$) indicates that the corresponding
states have $z$-components of their dipoles that are all opposite to
those of state $i$ (for instance, state $\beta$$'$ has $z$-components
of the dipoles being positive in the medium while being negative in
the wires, as exactly opposite to state $\beta$).  During this path
from state 2$'$ to state 2, the gyrotropic
coefficients $g_{11}$ and $g_{22}$ can be negative (from state 2$'$ to state 3$'$) or
positive (from state 5$'$ to state 2), depending on the sign of the
polarization.

Such possibility of having both negative and positive
gyrotropic coefficients in the same system originates from the fact
that the polarization can be down or up, and is consistent with
Eqs. (\ref{eq38}), (\ref{eq35a}) and (\ref{eq33}). As a result, one
can turn the polarization of light either in clockwise or
anticlockwise manner in electrotoroidic systems, via the control of
the direction of the polarization by an external electric field --
which induces the switching between the dextrorotatory and
laevorotatory forms of these materials (see states 1 and 1$'$). Such
control may be promising for the design of original devices \cite{footnotedegen,footnoteconj}.

Figure 3 shows how the gyrotropic coefficient $g_{11}$  depends on temperature. One can clearly see that $g_{11}$ significantly increases as the temperature increases up  to 240K. As indicated in the figure, the temperature behavior of $g_{11}$ is very well fitted by
$A/\sqrt{(T_C-T)(T_G-T)}$, where $A$ is a constant, $T_C=240K$ is the lowest temperature at which the polarization vanishes and $T_G=330K$ is the lowest temperature at which  the electric toroidal moment is annihilated  \cite{submitted}.
In order to understand such fitting,  let us combine Eqs (\ref{eq33}), (\ref{eq35a}) and
(\ref{eq38})  for the studied case, that is:
 \begin{equation}
\label{eqnew}
g_{11} =-\frac{4\pi \omega }{c}\frac{dG_3 }{dE_3
}=
-\frac{4\pi \omega }{c} \chi^{(G)}_{3n}\frac{\partial h_n}{\partial
P_l}\chi^{(P)}_{l3}=-\frac{4\pi \omega }{c} (\lambda_{n3l3} +\lambda _{n33l}+\lambda_{3nl3} +\lambda _{3n3l}) \chi^{(G)}_{3n} G_3 P_3
 \chi^{(P)}_{l3}
\end{equation}

The usual temperature dependencies of the order parameter and its conjugate field  imply that $G_3$ and $P_3$ should be proportional to $\sqrt{(T_G-T)}$ and $\sqrt{(T_C-T)}$, respectively, while their responses,
$ \chi^{(G)}_{3n}$ and  $\chi^{(P)}_{l3}$, should be proportional to $1/(T_G-T)$ and $1/(T_C-T)$, respectively. This explains why  the behavior of $g_{11}$ as a function of $T$ is well described by $A/\sqrt{(T_C-T)(T_G-T)}$.

\section{Summary}

In summary, we propose the existence of a new class of spontaneously optically active materials, via the use of different techniques (namely, analytical derivations, phenomenologies and first-principles-based simulations). These materials are electrotoroidics for which the electric toroidal moment changes  linearly under an applied electric field. Such linear change is demonstrated to occur if at least one of the following two conditions is satisfied: (1) the electric toroidal moment is coupled to a spontaneous
electrical polarization; or (2) the electric toroidal moment is coupled to strain and the whole system
is piezoelectric. We also report a realization of case (1), and further show that applying an electric field in such a case allows a systematic control of the  sign of the optical rotation,  via  a field-induced transition between the dextrorotatory and laevorotatory forms. We therefore hope that our study deepens the current knowledge of natural optical activity and will be put in use to develop novel technologies.

This work is financially supported by ONR Grants N00014-11-1-0384 and
N00014-08-1-0915 (S.P. and L.B.), ARO Grant W911NF-12-1-0085 (Z.G. and L.B.), NSF grant DMR-1066158 (L.L. and L.B.).  I.S acknowledges support by
Grant MAT2012-33720 from the Spanish Ministerio de Econom\'{\i}a y
Competitividad. S.P. appreciates Grant 12-08-00887-a of Russian Foundation for Basic Research. L.B. also acknowledges discussion with scientists sponsored
by  the Department of Energy, Office of Basic Energy
Sciences, under contract ER-46612, Javier
Junquera and Surendra Singh.




\begin{figure}
\centering
\resizebox{0.9\textwidth}{!}{\includegraphics{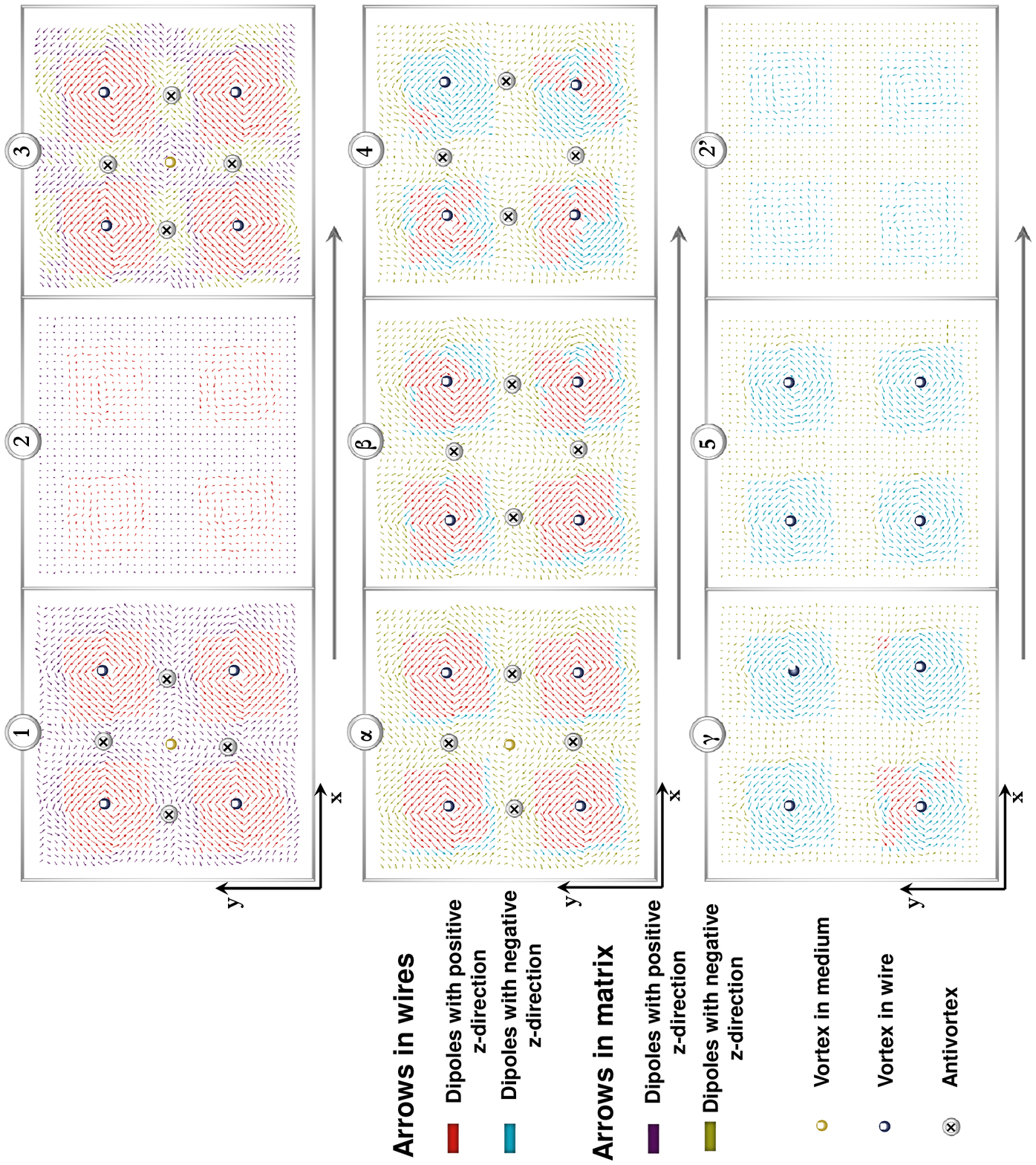}}
\caption{Dipole arrangement in the (x,y) plane of the studied nanocomposite for the states playing a key role in the occurrence of gyrotropy. The four wires are made of pure BaTiO$_3$, and the medium is mimicked to be formed by BST solid solutions having a 85\% Sr composition. See text
for the labels and meanings of the different panels.} \label{F1}
\end{figure}

\begin{figure}
\centering
\resizebox{0.9\textwidth}{!}{\includegraphics{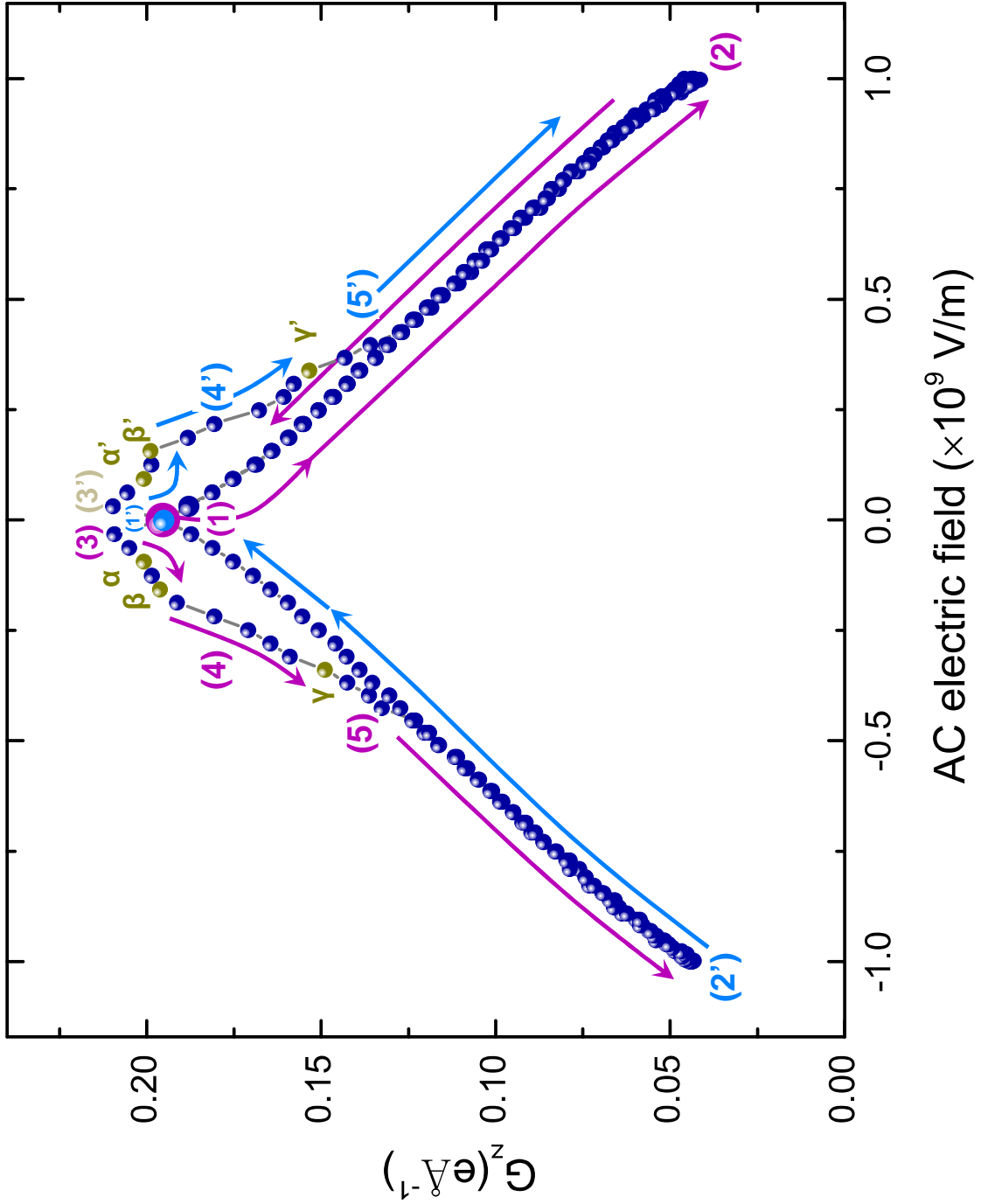}}
\caption{Predicted hysteresis loops in the studied nanocomposite at 15K, for a frequency of 1GHz. Panel (a) and (b) show the electrical toroidal moment and polarization, respectively,  as a function of the value of the $ac$ electric field. In these panels, the number and symbols inside parenthesis refer to the states displayed in FIG. 1.} \label{F2}
\end{figure}

\begin{figure}
\centering
\resizebox{0.9\textwidth}{!}{\includegraphics{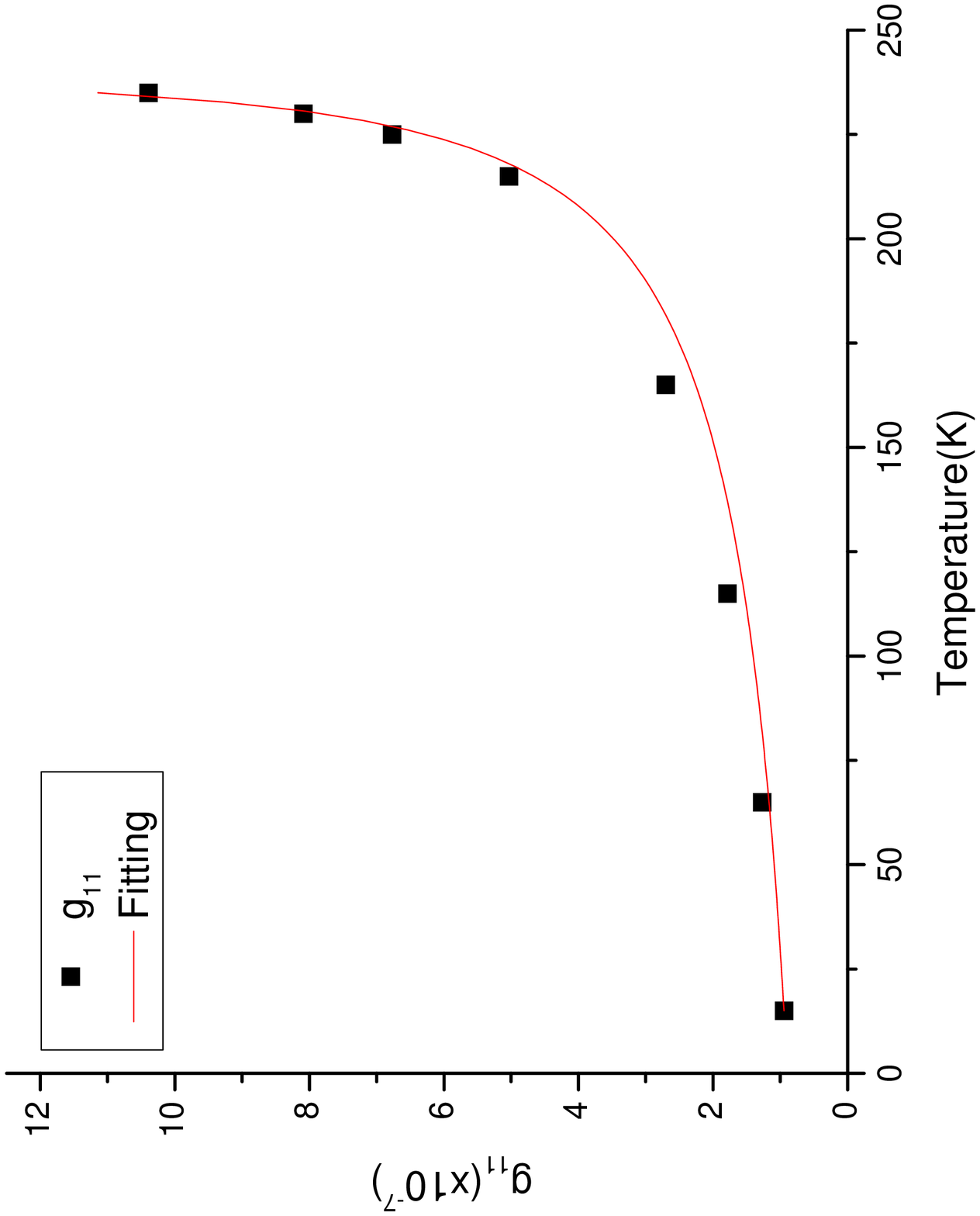}}
\caption{Temperature behavior of the  $g_{11}$ gyrotropic coefficient in the nanocomposite studied in the manuscript. The solid lines represent the fit of the data by $A/\sqrt{(T_C-T)(T_G-T)}$.} \label{3}
\end{figure}


\begin{thebibliography}{1}
\bibitem{footnoteconj} Note that the homogeneous electric field is the field conjugate of the electrical polarization but is not the field conjugate of the electrical toroidal moment. As a result (and as proven by our simulations), applying an electric field can change the direction of the polarization but can not change the direction of the electrical toroidal moment. This explains why an electric field can control the chirality and optical activity in  electrotoroidic systems.
\end{thebibliography}
\end{document}